\documentclass[%
 reprint,
superscriptaddress,
 amsmath,amssymb,
 aps,
floatfix,
]{revtex4-2}

\usepackage{graphicx}% Include figure files
\usepackage{dcolumn}% Align table columns on decimal point
\usepackage{bm}% bold math
\usepackage{xcolor}% text-color
\usepackage{hyperref}% add hypertext capabilities
\usepackage{multirow}
\usepackage[utf8]{inputenc}
\usepackage[T1]{fontenc}
\usepackage{mathrsfs}
\usepackage{CJK}
\CJKnospace
\usepackage{txfonts}
\usepackage[normalem]{ulem}
\usepackage{tikz}
\usepackage{soul}

\begin{document}
\begin{CJK}{UTF8}{mj}

%\preprint{APS/123-QED}

\title{
Global decomposition of networks into multiple cores formed by local hubs
}% Force line breaks with \\

\author{Wonhee Jeong (정원희)}
\thanks{Present address: %Center for Complex Networks and Systems Research, Luddy School of Informatics, Computing, and Engineering, Indiana University, Bloomington, IN, USA
Luddy School of Informatics, Computing and Engineering, Indiana University, Bloomington, Indiana 47408, USA}
\affiliation{%
    The Research Institute of Natural Science, Gyeongsang National University, Jinju 52828, Korea
}%

\author{Unjong Yu (유운종)}
\email[Contact author: ]{uyu@gist.ac.kr}
\affiliation{%
   Department of Physics and Photon Science \& Research Center for Photon Science Technology, Gwangju Institute of Science and Technology, Gwangju 61005, Korea
}%

\author{Sang Hoon Lee (이상훈)}
\email[Contact author: ]{lshlj82@gnu.ac.kr}
\affiliation{%
    The Research Institute of Natural Science, Gyeongsang National University, Jinju 52828, Korea
}%
\affiliation{%
    Department of Physics, Gyeongsang National University, Jinju 52828, Korea
}%
\affiliation{%
Future Convergence Technology Research Institute, Gyeongsang National University, Jinju 52849, South Korea
}%

\date{\today}% It is always \today, today,

\begin{abstract}
Networks are ubiquitous in various fields, representing systems where nodes and their interconnections constitute their intricate structures. We introduce a network decomposition scheme to reveal multiscale core-periphery structures lurking inside, using the concept of locally defined nodal hub centrality and edge-pruning techniques built upon it. We demonstrate that the hub-centrality-based edge pruning reveals a series of breaking points in network decomposition, which effectively separates a network into its backbone and shell structures. Our local-edge decomposition method iteratively identifies and removes locally least connected nodes, and uncovers an onion-like hierarchical structure as a result. Compared with the conventional $k$-core decomposition method, our method based on relative information residing in local structures exhibits a clear advantage in terms of discovering locally crucial substructures. As an application of the method, we present a scheme to detect multiple core-periphery structures and the decomposition of coarse-grained supernode networks, by combining the method with the network community detection. 
\end{abstract}


\maketitle

%\tableofcontents

\section{Introduction}\label{sec:level1}\protect

Networks have attracted considerable attention across various fields due to their omnipresence in nature and society~\cite{Newman2018,Barabasi2016,MenczerFortunatoDavis2020}. Consisting of nodes for individual objects of our interest and edges connecting them, a network succinctly represents an interacting system, which is a quintessential topic of statistical mechanics~\cite{Dorogovtsev2022}. Compared with traditional topics of statistical mechanics, perhaps a notably distinct feature of the recently developed theory of highly heterogeneous or ``complex'' networks is the existence of a few dominant elements that can govern the entire system. The identification of such important nodes and their disproportionately significant influence have been studied extensively~\cite{Newman2018,Barabasi2016,MenczerFortunatoDavis2020,Dorogovtsev2022}. One noticeable example in terms of both popularity and significance is the number of neighboring nodes for each node, which is called the degree. It affects a number of key aspects of networks, e.g., the robustness under failures and attacks~\cite{Albert2000,Attack_Holme,Jeong_2022}, the epidemic spreading~\cite{Pastor-Satorras2001}, critical phenomena~\cite{Dorogovtsev2008}, the controllability~\cite{Liu2011}, etc.

Roughly, there are two streaks of research for identifying important nodes with large degrees within a network: one focuses on individual nodes, e.g., by detecting the ones with large degrees or many connections to the rest of the network, usually dubbed as ``hubs''~\cite{Barabasi1999}, and the other identifies a group of important nodes, e.g., by detecting ``core'' nodes that are well connected to both each other inside and outside the group ~\cite{kcore_kitsak,kcore,Borgatti2000,Holme2005,Csermely2013,Rombach2017,Tang_CCF_cp,Kojaku_PRE_cp,Kojaku_NJP_cp,Yang_ESA_cp,Gallagher_SCIADV_cp}.
In contrast to the simplest concept of hub nodes by \emph{globally} counting their neighbors, when it comes to the core node groups, individual nodes' degree \emph{relative} to their peers in the group is also crucial. The latter is precisely captured by the measure called ``hub centrality'' introduced in the series of previous works~\cite{Jeong_2021,Jeong_2022} by some of the authors of this paper, to detect such \emph{local hubs} in the context of game theory and cascading failure. The hub centrality, defined as the normalized rank of each node within the node group composed of the node itself and its neighbors in terms of degree, successfully identifies locally important nodes. Both global and local hubs play profound roles across various dynamical systems, such as cascading failure~\cite{Jeong_2022,cascading_Valdez,cascading_Buldyrev}, disease spreading~\cite{kcore_kitsak}, vaccination~\cite{vaccination_Cohen}, and evolutionary game theory~\cite{cooperation_Gomez}, 
according to their unique characteristics.

In this regard, we point out that most of conventional methods to detect 
core nodes in networks~\cite{kcore_kitsak,kcore,Borgatti2000,Holme2005,Csermely2013,Rombach2017,Tang_CCF_cp,Kojaku_PRE_cp,Kojaku_NJP_cp,Yang_ESA_cp,Gallagher_SCIADV_cp} almost exclusively utilizes the concept of global connections in terms of degree, without enough consideration of nodes' relative position within their neighbor groups. Well-known examples include the decomposition of networks based on degrees, or the \textit{k}-core decomposition, which iteratively removes nodes with fewer than \textit{k} connections until only nodes with at least \textit{k} connections remain~\cite{kcore_kitsak,kcore}. Another related decomposition is the identification of the core-periphery structure~\cite{Borgatti2000,Holme2005,Csermely2013,Rombach2017,Tang_CCF_cp,Kojaku_PRE_cp,Kojaku_NJP_cp,Yang_ESA_cp,Gallagher_SCIADV_cp} by detecting the core nodes with statistically dense connections to the entire network and treating the rest as periphery. For all of these approaches, one simply takes a degree of a node as a face value without consideration of its aforementioned relative position. When there are a variety of local groups in heterogeneous sizes, however, as many real-world networks would actually be so, a particular degree value can make a locally strong hub that governs the entire dynamical property near the hub belonging to a small group, while the same degree value may correspond to a mediocre node belonging to a much larger group. In other words, the mixture of heterogeneous degree distributions~\cite{Barabasi2016} and heterogeneous locally dense structures or ``communities''~\cite{Porter2009,Fortunato_review,Fortunato2022} requires a meticulous approach to decomposing a network using nodes' degree as a tool. 

By discriminating the differential effects of the global and local hubs or cores, it would be possible to significantly enhance our understanding of both structural and dynamical aspects of networks. 
In this paper, we introduce the local-edge decomposition, which is based on the product of hub centrality values of the nodes connected by the edge of interest. One particularly promising aspect of this hub-centrality-based edge-pruning process is the presence of a natural breaking point between the zero and nonzero edge-importance values, detected by the giant component size, and we take the series of such breaking or cusp points to build a systematic procedure to decompose a network.
It assigns nodes' hierarchical levels and uncovers the onion structure of networks~\cite{ZXWu2011}, as we demonstrate for real and model networks. 

Because we use the concept of local hubs, our decomposition scheme takes a unique viewpoint of putting local hubs at the highest hierarchical level, and this perspective will open new possibilities for applications. 
Among the possibilities, we study and propose the core-periphery structure and the score function to find it, based on the nodes' local-hub-based hierarchical levels.  
We finalize the paper by extending our method by both zooming-in and zooming-out---core-periphery structures within communities and coarse-graind supernode networks, which may provide a crucial clue to solve the conundrum of the necessity for ``something else'' in core-periphery~\cite{Kojaku_NJP_cp} by finding multiple cores composed of local hubs.

\section{Decomposition of Networks}\label{sec:level2}\protect

The notion of network decomposition in this study refers to the process of extracting the most essential part of a network by iteratively peeling the relatively less important parts~\cite{kcore_kitsak,kcore}.
It allows for the identification and focused study of the most critical parts of the network, such as key infrastructure nodes in a power grid that, if failed, could cause widespread outages~\cite{YYang2017}, or influential individuals in a social network who can significantly impact the spread of information or diseases~\cite{Kempe2003}.
This targeted approach facilitates a deeper understanding and more detailed analysis of key components, ultimately leading to improved network performance and resilience.
In this section, we provide the result of edge pruning based on different criteria, introduce a local-hub-based strategy as our main scheme, and compare it with the conventional \textit{k}-core decomposition based on global degree values~\cite{kcore_kitsak,kcore}.

\subsection{Edge pruning and cusp point}\label{sec:level2-1}\protect

\begin{figure*}[tb]
\centering
\includegraphics[angle=270,width=1.7\columnwidth]{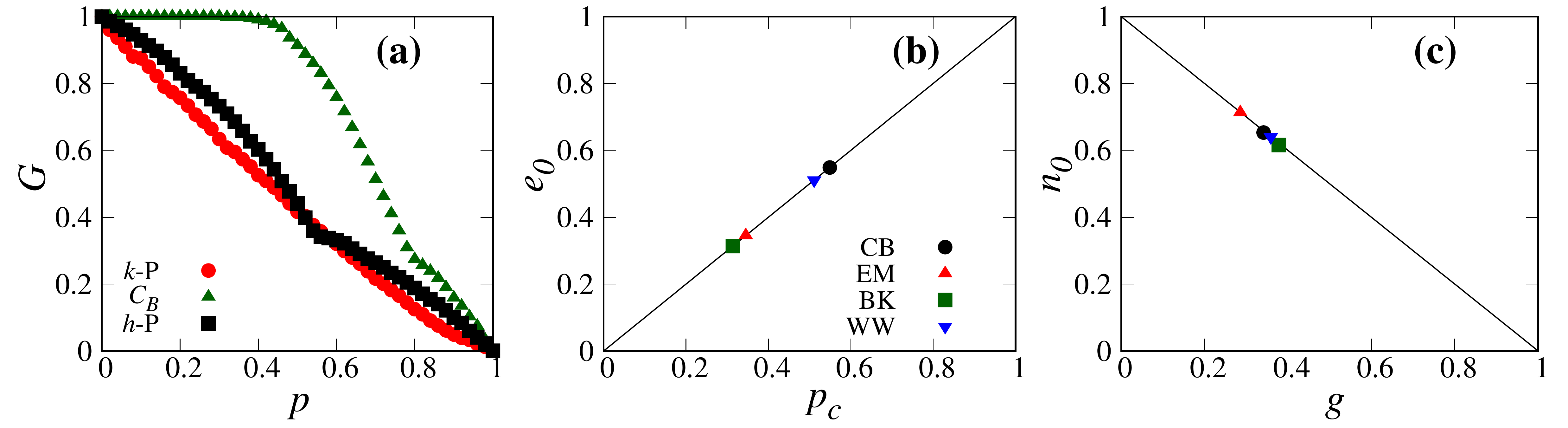}
\caption{(a) Relative size $G$ of the giant component as a function of the fraction $p$ of removed edges. The network is a collaboration network in the field of computational geometry. (b)--(c) Fraction $e_0$ of edges with $\mathscr{E}^{h\mbox{-}\mathrm{P}}=0$ and the fraction $n_0$ of nodes with $h=0$ as a function of the cusp point $p_c$ and the giant component size $g$ with zero hub-centrality nodes removed, respectively. Each point represents the outcome of each individual network listed in Table~\ref{tab1}. The solid lines in (b) and (c) represent $e_0 = p_c$ and $n_0 = 1-g$, respectively.}
\label{fig1}
\end{figure*}

To quantify edges' importance to set the criterion for decomposition, we try edge betweenness centrality related to the shortest-path-based global transport dynamics~\cite{Freeman1977} and the product of nodes' importance, which are attached to both ends of the edge.
In the latter case, the property of an edge that connects nodes $i$ and $j$ is given by the product of values assigned to each node as
\begin{eqnarray}
\mathscr{E}_{ij} = \psi_i \psi_j \,,
\label{Eproduct}
\end{eqnarray}
where $\psi_i$ is a certain property of node $i$.
In this paper, we try the degree representing the global connectivity and the hub centrality from the normalized local rank in connectivity~\cite{Jeong_2021,Jeong_2022} proposed by some of the authors of this paper.
The hub centrality of a node is the fraction of its neighbors with strictly lower degrees than the node itself~\cite{Jeong_2021,Jeong_2022}, given by
\begin{eqnarray}
h_i = \frac{N_{<k_i}}{k_i} \,,
\label{hc}
\end{eqnarray}
where $N_{<k_i}$ is the number of neighbors of node $i$ with lower degrees than the node $i$ itself, and the denominator $k_i$ is the degree of the node $i$ for normalization $0 \le h_i \le 1$.
We first try examining the effects of three types of edge importance for pruning edges in network decomposition, starting from widely used centralities: degree and betweenness centrality, in comparison to the hub centrality. We use
the edge betweenness centrality ($C_B$)~\cite{Newman2018,Barabasi2016,MenczerFortunatoDavis2020} by directly setting $\mathscr{E}^{C_B}_{ij} \equiv C_B(i,j)$, the degree product ($k\mbox{-}\mathrm{P}$) by setting $\mathscr{E}^{k\mbox{-}\mathrm{P}}_{ij} = \psi_i \psi_j$ where $\psi_i \equiv k_i$, and hub-centrality product ($h\mbox{-}\mathrm{P}$) by setting $\mathscr{E}^{h\mbox{-}\mathrm{P}}_{ij} = \psi_i \psi_j$ where $\psi_i \equiv h_i$.

In our edge-pruning strategy, we calculate the set of edge importance $\{ \mathscr{E}_{ij} \}$ in the original network and use it throughout the process until the end. We do not recalculate the hub centrality and its product during the edge removal process, as it is not our main interest to investigate the modified structure itself~\footnote{Another side effect of such recalculation is that the whole process would depend heavily on the order of removed edges with the same value of $\mathscr{E}_{ij}$, which leads to much more complicated situations in practice that requires ensemble average over multiple realizations.}; rather, we extract more central parts in terms of the edge importance in the original structure as we proceed.
We implement repeated edge pruning~\footnote{We suggest using the heap queue structure in numerical simulations for this.}, which removes the least important edges at each time step. We first examine a natural measure to characterize the edge-pruning process, which is the relative size $G = |\mathcal{G}|/N$ of giant (i.e., the largest connected) component $\mathcal{G}$ with respect to the original network size $N$, as a function of the fraction $p$ of removed edges. To focus on the fragmentation caused by the edge-pruning process, for all of our numerical studies, we use the giant component of the original networks (the sizes of which are listed inside the parentheses in Table~\ref{tab1}) in the beginning. 

\begin{table*}[tb]
\caption{Information of real-world networks considered in the work: number of nodes ($N$), the size ($|\mathcal{G}|$) of the giant component $\mathcal{G}$, number of edges ($E$), number of edges of the giant component ($E_G$), average degree ($\langle k \rangle$), average degree of the giant component ($\langle k \rangle_G$), average local clustering coefficient ($c_l$), and assortativity ($r$). When implementing local-edge decomposition, it is conducted within the giant component $G$. The collaboration network is the authors' collaboration network in the field of computational geometry.}
\begin{ruledtabular}
\begin{tabular}{lllllll}
%\hline
Networks & $N$ ($|\mathcal{G}|$) & $E$ ($E_G$) & $\langle k \rangle$ ($\langle k \rangle_G$) & $c_l$ & $r$ & Ref. \\ \hline
Collaboration (CB) & $6\,158$ ($3\,621$) & $11\,898$ ($9\,461$) & $3.864$ ($5.225$) & $0.485$ & $0.242$ & \cite{comgeo_net} \\
Email (EM) & $33\,696$ ($33\,696$) & $180\,811$ ($180\,811)$ & $10.731$ ($10.731$) & $0.509$ & $-0.116$ & \cite{email_net} \\
Brightkite (BK) & $58\,228$ ($56\,739$) & $214\,078$ ($212\,945$) & $7.353$ ($7.506$) & $0.172$ & $0.010$ & \cite{brightkite_net} \\
Wikipedia word (WW) & $146\,005$ ($145\,145$) & $656\,999$ ($656\,230$) & $8.999$ ($9.042$) & $0.602$ & $-0.062$ & \cite{WordNet_net} \\
%\hline
\end{tabular}
\end{ruledtabular}
\label{tab1}
\end{table*}

A typical example is the case of the collaboration network in the field of computational geometry (CB)~\cite{comgeo_net} (see Table~\ref{tab1} for basic statistics) shown in Fig.~\ref{fig1}(a); qualitatively similar results are observed across other networks listed in Table~\ref{tab1}, along with those from other networks not listed there.
In the case of $k\mbox{-}\mathrm{P}$, $G$ decreases monotonically.
For $C_B$, the value of $G$ remains high when $p \lesssim 0.4$, but $G$ decreases rapidly once $p \gtrsim 0.4$.
Notably, there is a cusp-like point at $p \approx 0.8$, which is the point where the rate of change in $G(p)$ qualitatively changes.
For $h\mbox{-}\mathrm{P}$, it shows a more prominent cusp point at $p = p_c \approx 0.5$, and the slope of $G(p)$ suddenly drops almost to zero when $p$ crosses the value $p_c$.
As we will discuss later, in the case of $h\mbox{-}\mathrm{P}$, a significant structural change in the network occurs at the cusp point, whereas we do not observe such a significant change for $C_B$.
Other networks listed in Table~\ref{tab1} also have the clear-cut cusp point when the edges are removed by $h\mbox{-}\mathrm{P}$ and show qualitatively similar behavior to Fig.~\ref{fig1}(a).
This seemingly puzzling behavior is, in fact, simply explained, with the hint from the observation of the fraction $e_0$ of edges with exactly null importance, i.e., the edges with $\mathscr{E}^{h\mbox{-}\mathrm{P}} =0$, which are precisely the edges connecting at least one node with the locally smallest degree: $h_i=0$ by the product rule.
As shown in Fig.~\ref{fig1}(b), for all of the four real networks we examine, the $e_0$ value coincides with the cusp point $p_c$ for each network. This implies that the transition between $p < p_c$ and $p > p_c$ corresponds to the part where the edges with $\mathscr{E}^{h\mbox{-}\mathrm{P}} = 0$ are removed (by our edge-pruning rule, they must be removed first) versus the part where the rest of the edges with $\mathscr{E}^{h\mbox{-}\mathrm{P}} > 0$ are removed. Note that we have tried other types of centrality measures including closeness centrality, but none of them shows this peculiar behavior.

Moreover, for any node $i$ with $h_i = 0$, by definition, all of its neighboring nodes should have larger degrees than $k_i$, so $\mathscr{E}^{h\mbox{-}\mathrm{P}}_{ij} = 0$ for any node $j$ connected to $i$. As a result, all of the edges connected to $i$ will be removed during the $p < p_c$ pruning process. Conversely, as already stated, each of the edge-removal processes for $p < p_c$ involves such a node with $h_i=0$. Therefore, the edge-pruning process for $p < p_c$ exactly corresponds to the process of pinpointing the nodes with $h_i = 0$ and removing all of their edges. This surgical removal effectively isolates those locally least connected nodes and leaves the rest of the network as a new giant component precisely at $p = p_c$. In principle, it is possible for a node with nonzero hub centrality to be isolated as a result of the removal of all of its neighbors with $h = 0$ even for $p < p_c$~\footnote{One can imagine a following (extreme) case: assume a node with $h>0$ connected to a number of other nodes, \emph{all of which} have $h=0$ and are connected to very large and dense clique-like subnetworks on the other side than the original node itself. In that case, by the elimination of the neighboring nodes, the original node will be isolated. This effect is reflected in the actual appearance of $g \equiv G(p = p_c) < 1 - n_0$ in Fig.~\ref{fig2}(d), but actually it is quite rare (usually less than $0.1\%$ of nodes in our empirical networks).}. However, one can check that the relation $g \equiv G(p = p_c) = 1 - n_0$ holds (under the assumption of the absence of such cases; in general, $g < 1 - n_0$ caused by the possibility of the aforementioned ``casualty'' node with $h>0$) for our empirical networks, where $n_0$ is the fraction of the nodes with $h_i = 0$ and $g$ is the fraction of the giant component remaining at $p = p_c$, from Fig.~\ref{fig1}(c). In other words, the separation of zero-hub-centrality nodes does not cause the noticeable separation of other nodes with nonzero hub centrality from the giant component up to $p = p_c$, and most nodes with $h>0$ belong to the new giant component at $p = p_c$.

From these observations, we deduce that, except for peculiar substructures such as linear chains attached to the component~\footnote{Even in the case of a long linear chain, one can easily show that only a single node with $h>0$ right next to the dead end is removed per chain and all of the other nodes in the chain has the value $h=0$ anyway.}, a network composed of a single giant component harbors a one-step deeper-level giant component formed by positive hub-centrality nodes inside and zero hub-centrality nodes attached to it outside. Simply speaking, inside nodes maintain the connectivity even when outside nodes are removed.
In terms of transport property, if we choose two different zero hub-centrality nodes (outside nodes), as each of them has connections to nonzero hub-centrality nodes (inside nodes), there is at least a path only through nodes with $h>0$, which is reminiscent of the backup-pathway-based notion of core-periphery~\cite{SHLee2014,Cucuringu2016} and highlights the role of the giant component at $p = p_c$ as a structural and dynamical backbone. On that threshold, the network is divided into a backbone composed of nodes with $h > 0$ and a shell mostly composed of nodes with $h=0$.

\subsection{Local-edge decomposition and node hierarchy}\label{sec:level2-2}\protect

\begin{figure*}[tb]
\centering
\includegraphics[angle=270,width=2\columnwidth]{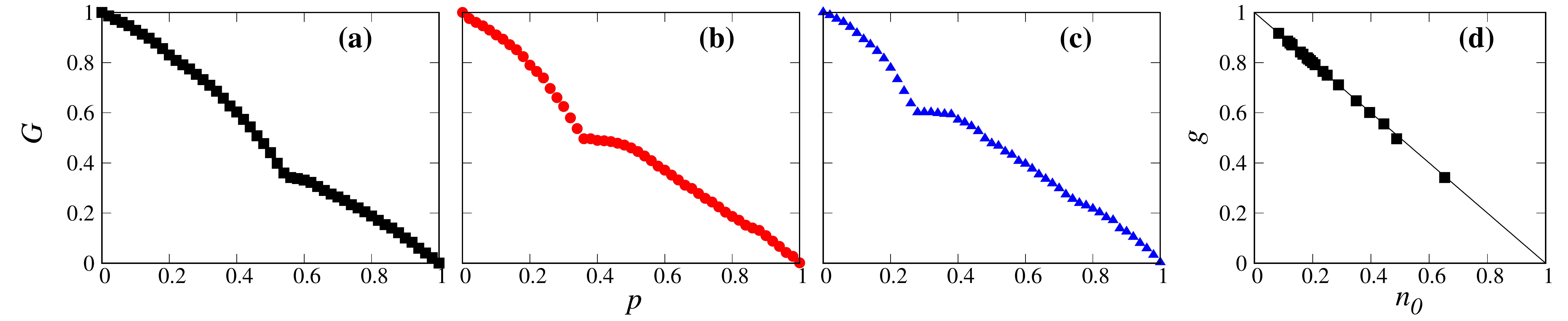}
\caption{(a)--(c) Relative size $G$ of the giant component as a function of the fraction $p$ of removed edges based on the hub-centrality product ($h\mbox{-}\mathrm{P}$). (a) The original collaboration network used in Fig.~\ref{fig1}. (b) Backbone of the original network. (c) Backbone of the original network's backbone. (d) Relative fraction $g$ of the giant component size with zero hub-centrality nodes removed as a function of the fraction $n_0$ of zero hub-centrality nodes in decomposed networks. The solid line represents $n_0=1-g$, which implies the set of zero hub-centrality nodes $\approx$ the shell at each decomposition level.}
\label{fig2}
\end{figure*}

The clear-cut separation between the backbone and the shell described in Sec.~\ref{sec:level2-1} provides us a nice natural cutoff to examine the locally important component. Then, why do we just stop there? We can use the process \emph{repeatedly}, by applying the same procedure to the backbone as a new network for decomposition, and so on. 
For the backbone [the giant component at the cusp point $p_c^{(1)}$ in Fig.~\ref{fig2}(a)], we recalculate the hub-centrality values for each node and implement the same edge-pruning process by the $h\mbox{-}\mathrm{P}$ rule.
Figure~\ref{fig2}(b) shows the results of such second-level edge-pruning, and we can observe a cusp point again, as in the original network Fig.~\ref{fig2}(a). Not surprisingly, it occurs as a number of nodes with $h > 0$ in the original network have now become nodes with $h = 0$ in the first-level backbone. 
Accordingly, the first-level backbone is once again separated into the second-level backbone and shell.
In the second-level backbone, which is the giant component at $p = p_c^{(2)}$ of Fig.~\ref{fig2}(b), we recalculate hub centrality and implement the edge-pruning process again.
As expected, it shows a cusp point at $p = p_c^{(3)}$ (the third level) as shown in Fig.~\ref{fig2}(c). 

We can continue this backbone-extraction process by removing the edges with $\mathscr{E}^{h\mbox{-}\mathrm{P}}=0$ at each level until all of the nodes are isolated. We refer to this process as the local-edge decomposition (LED). For the readers, we provide the following step-by-step guide for numerical simulation:
\begin{enumerate}
    \item \textbf{Initial Calculation}: Calculate the hub centrality of each node in the network (level $L = 0$).
    \item \textbf{Edge Pruning}: For each level $L$, identify and remove all of the edges where the product of hub centrality values is zero, i.e., $\mathscr{E}^{h\mbox{-}\mathrm{P}}=0$. 
    When there is no edge with $\mathscr{E}^{h\mbox{-}\mathrm{P}}=0$ remaining, the fraction $p$ of removed edges is equal to $p_c$ at level $L$ and the remaining giant component becomes the new network at level $(L+1)$.
    \item \textbf{Recalculation}: At the beginning of each new level $(L+1)$, recalculate the hub centrality for the new network at level $(L+1)$.
    \item \textbf{Iteration}: Repeat the edge-pruning and recalculation process described in 2 and 3 above by increasing level $L$.
    \item \textbf{Termination}: Continue this iterative process until no more edges can be removed (i.e., all nodes are isolated).
\end{enumerate}
Figure~2(d) demonstrates that our earlier argument holds well; $n_0 \approx 1 - g$ (for $p < p_c$) and the edges are eliminated if and only if $\mathscr{E}^{h\mbox{-}\mathrm{P}}=0$ at each level. As a result, all of the $(n_0, g)$ pairs at different levels fall on top of the $n_0 = 1 - g$ line.
The other networks listed in Table~\ref{tab1} also show qualitatively similar results to Fig.~\ref{fig2}.
The final level is composed of nodes with the same degree and the edges with $\mathscr{E}^{h\mbox{-}\mathrm{P}} = 0$ (no node's degree can be smaller than any other nodes' degree), so the whole process is naturally terminated by the elimination of the entire edges altogether along with all of the nodes; generally, it forms a clique in real-world networks according to our observation. Therefore, we conclude that a network can be decomposed as an onion-like structure~\cite{ZXWu2011}, wherein we extract the core of the onion by removing locally least connected nodes iteratively through LED.

\begin{figure}[tb]
\centering
\includegraphics[angle=0,width=0.8\columnwidth]{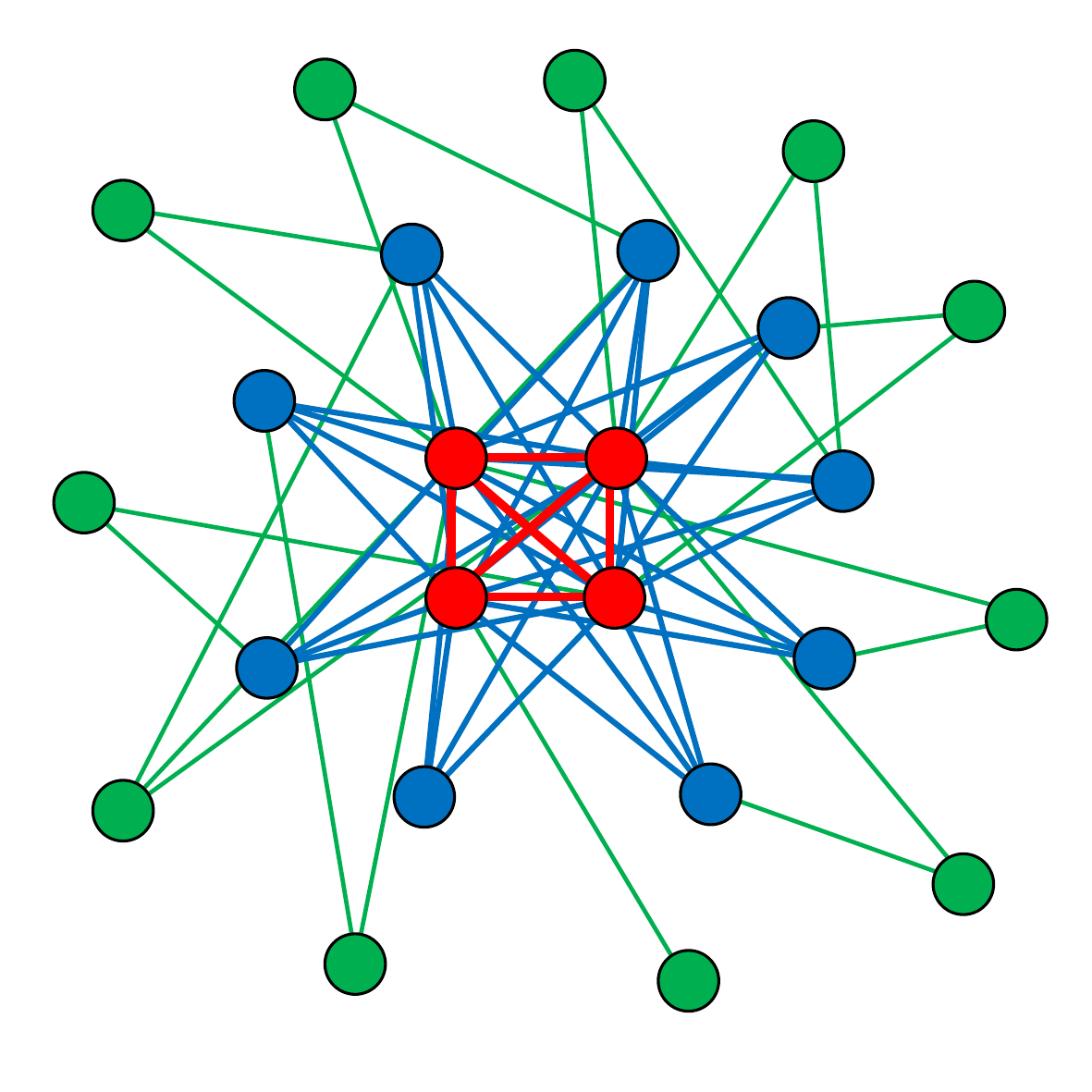}
\caption{An example network to illustrate the LED decomposition. The nodes filled with the same color belong to the same hierarchical level. The green, blue, and red nodes belong to the lowest, intermediate, and highest levels, respectively. The color and thickness of the edges indicate the decomposition levels as well.}
\label{fig3}
\end{figure}

As an illustration of this iterative LED, we show a simple example in Fig.~\ref{fig3}, where different colors (and edge thickness) represent the decomposition level. 
The green nodes have zero hub centrality and are connected to the green edges with $\mathscr{E}^{h\mbox{-}\mathrm{P}}=0$, and they are decomposed first.
At the decomposed network, which is composed of blue and red nodes, we recalculate hub centrality and remove the blue edges with $\mathscr{E}^{h\mbox{-}\mathrm{P}}=0$.
As a result, the blue nodes are decomposed, and only the red nodes that form a $4$-clique with the red edges remain, which will eventually be removed at the next level. In other words, the green, blue, and red nodes are hierarchically organized to constitute levels $0$, $1$, and $2$, respectively.

\subsection{Local-edge decomposition of the Barab{\'a}si-Albert model}\label{add1}\protect

\begin{figure}[tb]
\centering
\includegraphics[angle=270,width=1\columnwidth]{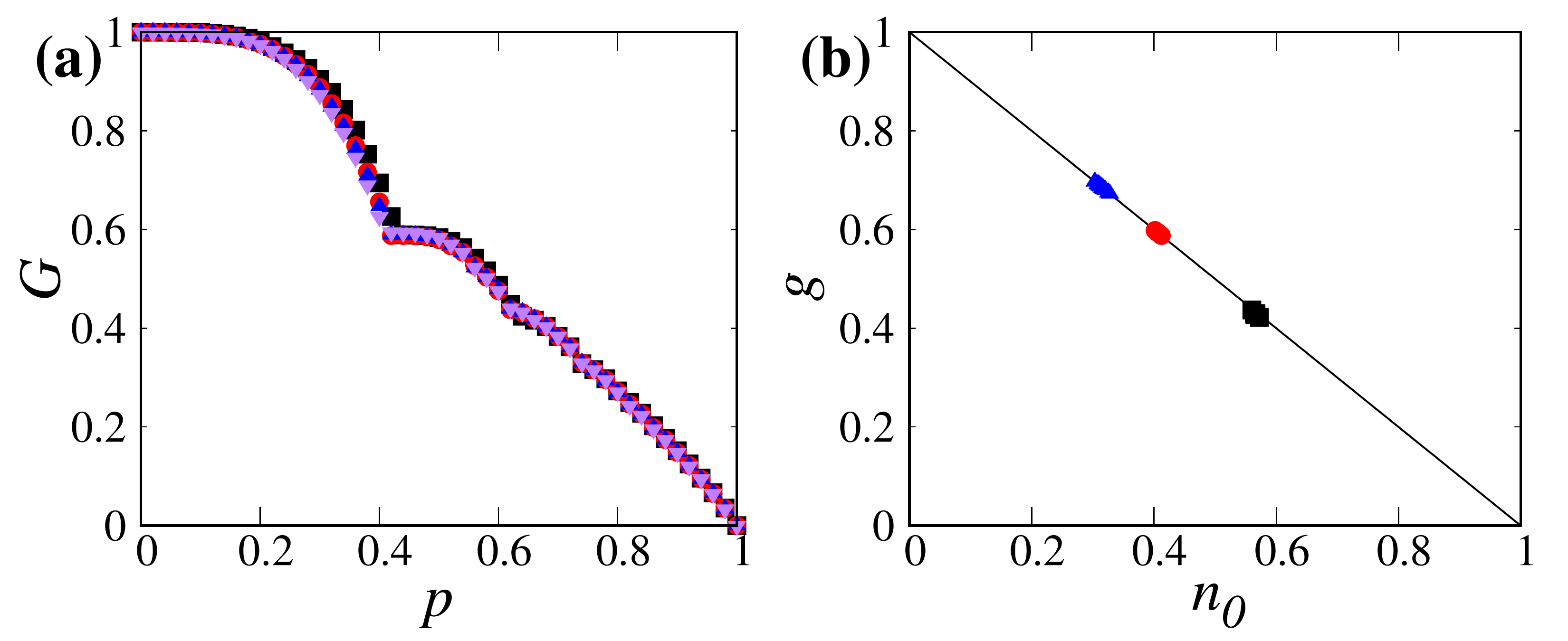}
\caption{%
(a) Relative size $G$ of the giant component as a function of the fraction $p$ of removed edges based on $h\mbox{-}\mathrm{P}$, applied to a single realization of the BA model~\cite{Barabasi1999} with $10^5$ nodes and the number $m = 4$ of stubs for each newly added node. The black squares indicate the original network. The red circles, blue triangles, and purple inverted triangles represent the results obtained for the primary, secondary, and tertiary backbones, respectively, of the original network.
(b) Relative fraction $g$ of the giant component size with zero hub-centrality nodes removed as a function of the fraction $n_0$ of zero hub-centrality nodes in decomposed networks at different levels. The black squares, red circles, and blue triangles represent the cases with $m = 2$, $4$, and $6$, respectively. We only plot the cases where the backbone with the number of nodes $> 1\%$ of that of the original network.
}%
\label{figBA}
\end{figure}

Our LED analysis of real networks may give the impression that it is not applicable to simple model networks without mesoscale heterogeneity. This assumption must be correct for a completely homogeneous mixture of random connections represented by the Erd\H{o}s-R{\'e}nyi (ER) random graph~\cite{Erdos1959} as all of the nodes there are topologically equivalent. In this subsection, however, we provide an example of a model network where its construction principle induces the sequential hierarchy of LED levels that is well-detected by our method. The historically important and celebrated Barab{\'a}si-Albert (BA) model~\cite{Barabasi1999} is well-known for its power-law degree distribution that gave birth to the famous notion of ``scale-free'' networks (SFNs). At the same time, however, it is relatively less emphasized that one of the key factors of the model, the growth by attaching nodes with a fixed number $m$ of stubs, provides the intrinsic correlation between individual nodes' time of inception and its degree~\cite{Jeong_2022,Barabasi1999a}, statistically speaking. As a result, in contrast to other SFN models without the ``growing'' mechanism~\cite{KIGoh2001,Caldarelli2002}, the BA model tends to be organized as in Fig.~\ref{fig3}, where central (peripheral) nodes correspond to older (newer) nodes.

Such an organizational structure is well-captured by the hub-centrality-based LED applied to the BA model, as shown in Fig.~\ref{figBA}, where we show the $G(p)$ and $g(n_0)$ curves for different LED levels as in the cases of real networks. The evidence cusp points in Fig.~\ref{figBA}(a) and the relation $g \approx 1 - n_0$ at each cusp point in Fig.~\ref{figBA}(b) indicate that our method successfully reveals the aforementioned organizational structure of the model. The most outstanding feature about the BA-model case is the fact that $G(p)$ curves for different levels are collapsed onto a single characteristic curve. This implies that in contrast to real networks [see Fig.~\ref{fig2}(a)--(c)] each LED level or backbone of the BA model has almost the same organizational structure starting from the original network, which is also reflected in the collapsed points $\left(n_0,g(n_0) \approx 1-n_0\right)$ for different levels at each $m$ value [Fig.~\ref{figBA}(b)]. This self-similar or scale-invariant property reminds us of more fundamental scale-free property than just the power-law degree distribution~\cite{Poggialini2024}. In retrospect, our edge-pruning processes correspond to the \emph{time-reversal processes} of the node and edge growing mechanism imposed in the creation of the BA model on average (the preferential attachment for creation versus the preferential detachment~\cite{MJLee2022} for LED), so the result is not too surprising. Moreover, even at a single level, in addition to the primary cusp point at $p \approx 0.4$, there are (less apparent) secondary ($p \approx 0.6$) and more cusp-like points in the BA model (again, at almost exactly the same $p$ for every level). The first cusp point is the point that the nodes with zero hub centrality are separated from the giant component as in real networks, but the others are causes by the collective and systematic decrement of the nodes with low hub-centrality values, which are absent in real networks without such an artificial systematic attaching mechanism of nodes with a fixed number of stubs.

\subsection{Comparison with the \textit{k}-core decomposition}\label{sec:level2-3}\protect

\begin{figure}[tb]
\centering
\includegraphics[angle=270,width=1\columnwidth]{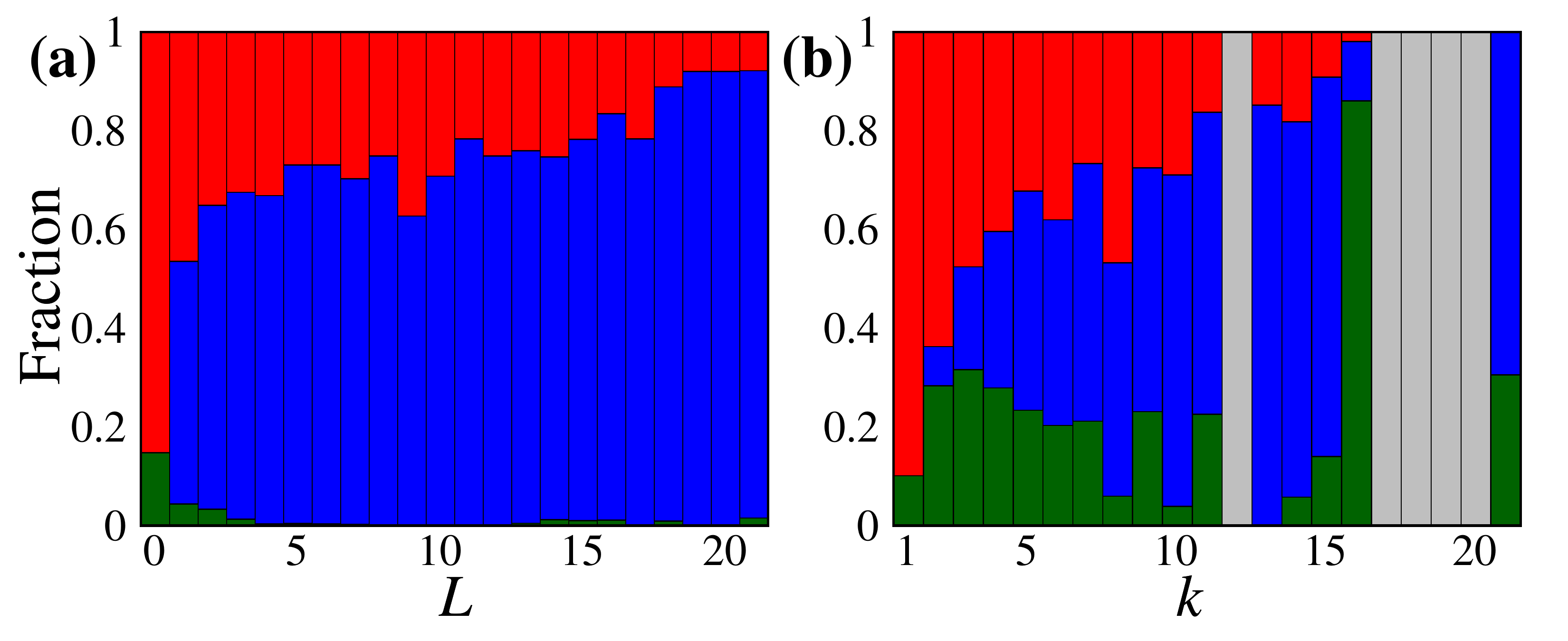}
\caption{%
Fraction of the level (or ``shell'' in the \textit{k}CD) relationship between nodes and their neighbors. The colors red, blue, and green indicate that nodes at a given level have neighbors with higher, lower, or the same level, respectively. The regions filled with gray represent the absence of nodes in the corresponding levels. Panel (a) is for the LED, and panel (b) is for the \textit{k}CD.
}%
\label{fig4}
\end{figure}

Our decomposition scheme will obviously remind anyone familiar with network science of the celebrated \textit{k}-core decomposition (\textit{k}CD)~\cite{FirstCourseBook,kcore_kitsak,kcore}. The \textit{k}CD is one of the early established methods to extract the most central part of a network, by iteratively removing the smaller-degree nodes.
Starting from $k = 1$, it peels out nodes with the minimum degree until no node has a degree smaller or equal to $k$ at each stage (the removed nodes for a given value of $k$ are called the $k$-shell, analogous to the ``level'' in our LED)~\footnote{In contrast to our decomposition method from the hub centrality, the degree values are changed in real-time and instantly reflected in the node-removal process.}, and continues this process by increasing $k$ until all of the nodes are removed; in this way, nodes are hierarchically decomposed as in our LED, with a similar final stage composed of a clique to ours. 

\begin{figure}
\includegraphics[width=\columnwidth]{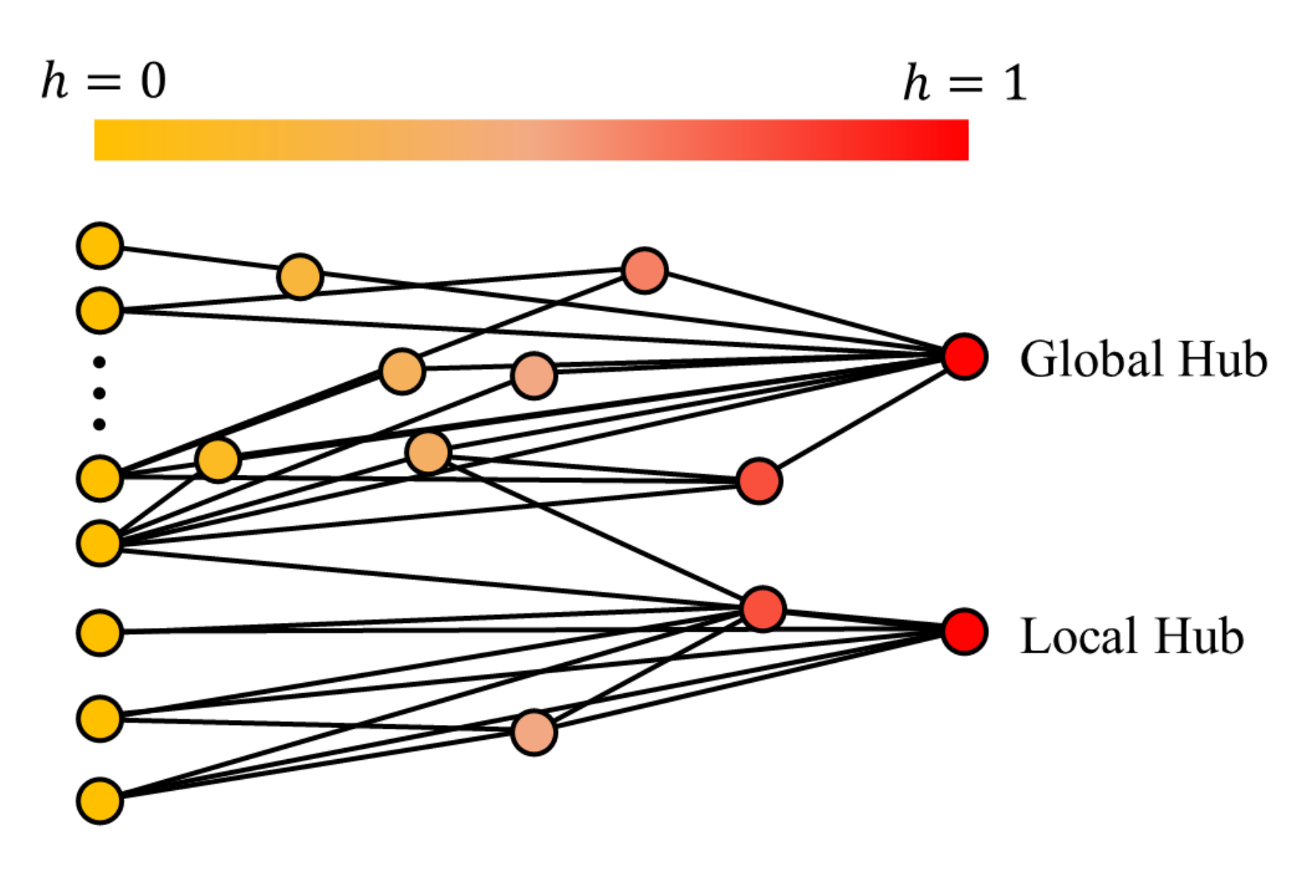}
\caption{
Schematic illustration of examples of global and local hubs, whose horizontal location is aligned with their hub-centrality values represented with the horizontal color bar with gradation.
}
\label{fig:local_hub}
\end{figure}

Despite the similarity, however, the crucial difference between the \textit{k}CD and our LED comes from the fact that the \textit{k}CD is based on degree (global information), and our LED is based on hub centrality (local information). One way to see the difference is to observe the inter-level connections. 
Figure~\ref{fig4} shows how nodes included in a given level are connected to other-level nodes for the CB network.
In the case of LED, as shown in Fig.~\ref{fig4}(a), the overall tendency indicates that higher-level nodes are connected to lower-level nodes naturally, but the connections between the same-level nodes are quite rare except for a few lowest levels. This happens because the same-level connections correspond to the edges connecting the two distinct zero hub-centrality nodes in a given level, which is only possible when the two nodes have exactly the same degree and both of them should have the lowest degree among their neighbors. In other words, the LED is a systematic way to find \emph{the edges essentially separating different hierarchical levels} of network organization. For \textit{k}CD, in contrast, there are a substantial number of connections between the same-level nodes as shown in Fig.~\ref{fig4}(b), because the nodes in the $k$-shell would just mean a similar edge density in the area regardless of the local degree gradient with respect to the neighbors. 

An illustrative way to see the stark difference between the LED and the \textit{k}CD is presented in Fig.~\ref{fig:local_hub}. In the LED, the highest-level nodes are \emph{local} hubs with $h \approx 1$ regardless of their degree values themselves, so the decomposition process gradually prunes edges \emph{simultaneously} in substructures (e.g., communities) with various different scales. This property will play a crucial role in dealing with multiple core-periphery structures later. In contrast, if the \textit{k}CD were used, the local hub would be removed much earlier than the global hub, so observing the local organizational structure with different scales would be much harder.
In addition, because the LED uses the local relative information, the nodes in a network are naturally composed of consecutive nonempty levels as shown in Fig.~\ref{fig4}(a). In contrast, for the  \textit{k}CD using the absolute degree values, it is possible for a certain $k$-shell to be empty as shown in the gray parts in Fig.~\ref{fig4}(b). As a result, for most cases, we obtain a more gradual decomposition of a network for the LED, compared with the \textit{k}CD. This is another advantage of using the LED, which provides more \emph{granular} information about network organization.

\section{Core-Periphery Structure of Networks}\label{sec:level3}\protect

The core-periphery structure (CP)~\cite{Borgatti2000,Holme2005,Csermely2013,Rombach2017,Tang_CCF_cp} is another fundamental mesoscale structure of networks regarding the gradually sparser or denser parts in a network; it implies that a network consists of a dense ``core'' and sparse ``periphery.''
%There have been a substantial number of methods to identify a single core within the network, but
Most early-day studies on the CP assume the existence of a single core and the periphery surrounding it understandably because it is simplest. However, recent studies started to acknowledge that for networks to have a nontrivial CP other than the one that can easily be separated by the degree values, it is essential to have a complicated CP composed of multiple cores (and not surprisingly, most real networks are ``complex'' enough to do so)~\cite{Kojaku_PRE_cp,Kojaku_NJP_cp,Yang_ESA_cp,Gallagher_SCIADV_cp}.
Considering the ubiquitous existence of community structures~\cite{Porter2009,Fortunato_review,Fortunato2022}, it is also reasonable to assume the presence of multiple cores. Our LED scheme is able to detect such structures, as we will present from now on.

\subsection{Core-periphery structure and score}\label{sec:level3-1}\protect

\begin{figure}[tb]
\centering
\includegraphics[angle=270,width=0.9\columnwidth]{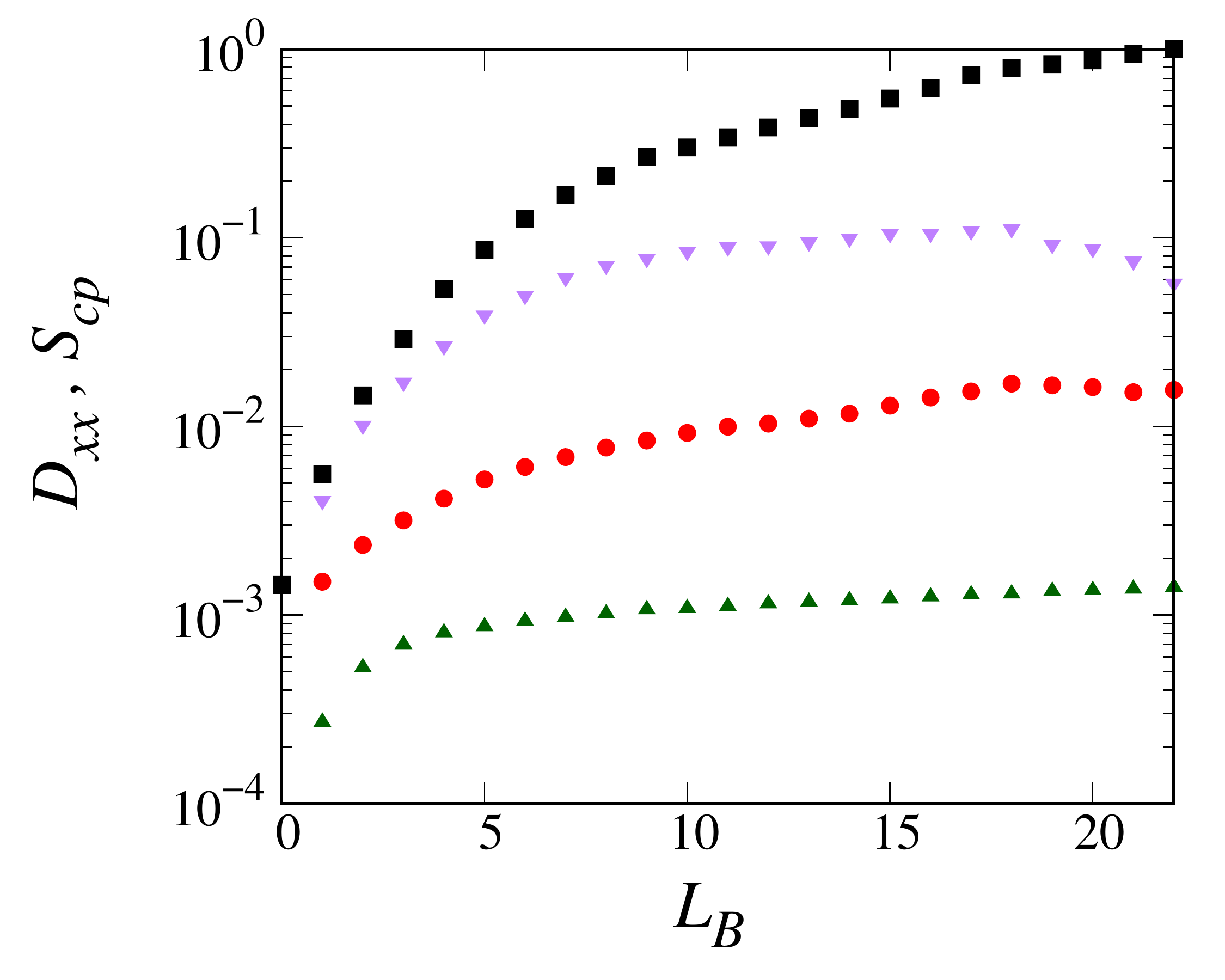}
\caption{
Connection densities of core-core, core-periphery, and periphery-periphery, along with the core-periphery score, as a function of the boundary level $L_B$. The black (square), red (circle), and green (triangle) points represent the connection densities of core-core, core-periphery, and periphery-periphery, respectively. The purple (inverse triangle) points indicate the core-periphery score. At $L_B=0$, the connection densities for core-periphery and periphery-periphery, as well as the core-periphery score, are zero. The core-periphery score reaches its maximum at $L_B=18$.
}
\label{fig5}
\end{figure}

First, to demonstrate that our LED identifies the single CP, we observe the connection density of core-core ($D_{cc} = 2M_{cc} / [N_c (N_c - 1)]$), core-periphery ($D_{cp} = M_{cp} / (N_c N_p)$), and periphery-periphery ($D_{pp} = 2M_{pp} / [N_p (N_p - 1)]$) edges, where $N_c$, $N_p$, $M_{cc}$, $M_{cp}$, and $M_{pp}$ correspond to the number of core nodes, that of periphery nodes, that of edges connecting core nodes, that of edges connecting core and periphery nodes, and that of edges connecting periphery nodes, respectively. The conventional notion of CP suggests the inequality~\cite{Borgatti2000,Holme2005,Csermely2013,Rombach2017,Tang_CCF_cp}
\begin{equation}
D_{cc}>D_{cp}>D_{pp} \,.
\label{eq:DccDcpDpp}
\end{equation}
We implement the LED in a network, classify nodes according to the hierarchical level $L$ as in Sec.~\ref{sec:level2-2}, and treat higher-level nodes as the core part. The key question is then to set the boundary between the core and the periphery, i.e., to propose a threshold value $L_B$, where the nodes in the levels $L \ge L_B$ and those in the levels $L < L_B$ are categorized as the core and periphery parts, respectively.
According to Fig.~\ref{fig5} for the CB network, for any value of $L_B$, the CP condition in Eq.~\eqref{eq:DccDcpDpp} is satisfied. As demonstrated in Sec.~\ref{sec:level2-2}, the nodes with the lowest relative degree in local neighbors are the first to be decomposed from the giant component, leading to higher connection densities among higher-level nodes compared with lower-level nodes.

To decide the most appropriate value of $L_B$ to accurately distinguish the CP, we introduce a core-periphery score $S_{cp}$ based on the following two conditions that we consider as the most ideal CP.
\begin{enumerate}
\item The core nodes are fully connected to each other.
\item All of the periphery nodes are connected to at least a core node, and no edges between periphery nodes exist.
\end{enumerate}
The condition $1$ is quantified with $D_{cc}$, and the condition $2$ is measured by $N_{pc} / N_p$, where $N_{pc}$ is the number of periphery nodes with at least one connection to core nodes, so that the ratio $N_{pc} / N_p$ represents the fraction of periphery nodes that are connected to the core nodes. In other words, the extent to which the core and periphery nodes satisfy the above two conditions is determined by $D_{cc}$ and $N_{pc}/N_p$, respectively.
To quantitatively assess the core-periphery structure using these criteria, we define the core-periphery score as,
\begin{eqnarray}
S_{cp} \equiv D_{cc} \frac{N_{pc}}{N_p} - D \, \mathrm{min}\{1,DN_c\} \,,
\label{score_cp}
\end{eqnarray}
where $D = 2(M_{cc} + M_{cp} + M_{pp})/[(N_c + N_p)(N_c + N_p - 1)]$ is the density of the entire network.
This represents the difference between the actual similarity and the null-model case, where the same overall connection density $D$ applies for all potential connections. The second term on the right-hand side of Eq.~\eqref{score_cp} is the expected value of the first term for the case of completely random connections (the uniform edge density $D$ throughout the entire network) without any prescribed division between core and periphery nodes, as $D_{cc} = D$ and $N_{pc}/N_p = \mathrm{min}\{1,DN_c\}$ in that case; if $D > 1/N_c$, for each periphery node there would be at least one edge to a core node on average, and if $D < 1/N_c$, the expected fraction of periphery node with at least one connection to a core node would be $D N_c$. 
The core-periphery score is maximized when both conditions are satisfied, thus it helps to identify the most optimal boundary level $L_B^*$ for distinguishing between the core and periphery of a network.
When a network exhibits an ideal CP structure, $S_{cp}$ is close to $1$.
In contrast, in a null model network, $S_{cp}=0$ because the core-periphery score represents the comparison of the network with the null model by definition.
As the boundary level $L_B$ increases from $1$ to $18$ ($S_{cp}$ for $L_B = 0$ is undefined), the distinction between core and periphery nodes becomes clearer, causing the core-periphery score to gradually increase.
At $L_B^* = 18$, the core-periphery score reaches its maximum value, and this means that nodes belonging to $L \ge L_B^* = 18$ form the core, while the remaining nodes form the periphery in this network.
However, when $L_B$ exceeds $L_B^* = 18$, $N_{pc}$ decreases and the core-periphery score starts to decrease.

In large-scale real networks, the assumption of single core-periphery separation has its clear limitation. Given that the macroscale network consists of various mesoscale structural features (multiple hubs, communities, etc.) and their interwoven mixture, the presence of multiple cores and a mesoscale core is an inherent characteristic of its structure.

\subsection{Core-periphery structure of communities and the supernode network}\label{sec:level3-2}\protect

\begin{figure*}[tb]
\centering
\includegraphics[angle=0,width=2\columnwidth]{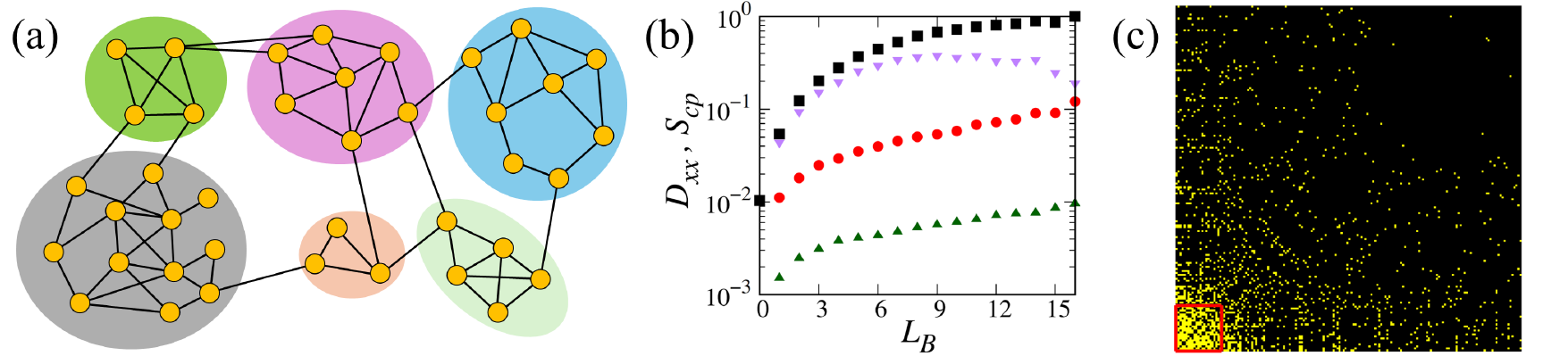}
\caption{%
(a) Example of a network with a community structure marked by different background colors. (b) Inside an arbitrary community of the CB network, we plot the connection densities of core-core, core-periphery, and periphery-periphery, along with the core-periphery score, as a function of the boundary level $L_B$. The same colors and shapes are used for the densities and the score as in Fig.~\ref{fig5}. The core-periphery score $S_{cp}$ reaches its maximum at $L_B=9$. (c) The adjacency matrix represents the connections (yellow in the black background) inside the community in panel (b), where nodes are sorted by their LED levels (higher levels toward the left bottom). The nodes with the lowest level are excluded for visibility, and the nodes inside the red square are the optimal set of core nodes of this community.
}%
\label{fig6}
\end{figure*}

\begin{figure*}[tb]
\centering
\includegraphics[angle=0,width=2\columnwidth]{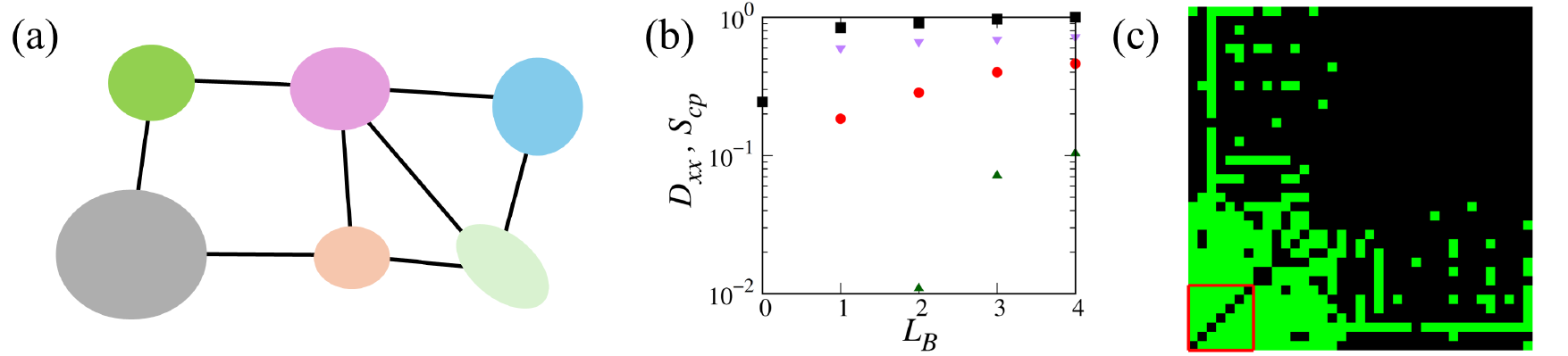}
\caption{%
(a) Supernode network, which is composed of the nodes corresponding to the communities in the network shown in Fig.~\ref{fig6}(a). (b) Connection densities of core-core, core-periphery, and periphery-periphery, along with the core-periphery score, as a function of the boundary level $L_B$ in the supernode network from the CB network. The same colors and shapes are used for the densities and the score as in Figs.~\ref{fig5} and \ref{fig6}. The core-periphery score $S_{cp}$ reaches its maximum at $L_B=4$. (c) The adjacency matrix represents the connections (green in the black background) between the supernodes in panel (b), where nodes are sorted by their LED levels (higher levels toward the left bottom). The supernodes inside the red square are the optimal core nodes of this community.
}%
\label{fig7}
\end{figure*}

In the previous subsection, we have found a single core of the network, but as introduced in the first part of Sec.~\ref{sec:level3}, more recent studies on the CP emphasize the necessity for the consideration of multiple CP in networks~\cite{Kojaku_PRE_cp,Kojaku_NJP_cp,Yang_ESA_cp,Gallagher_SCIADV_cp}. One hint from the literature is the fact that the mixture of community structures and CP is interchangeably expressed as ``CP inside communities'' and ``communities inside CP'' (see Fig.~1.1 of Ref.~\cite{Rombach2017}). In this subsection, we take both viewpoints by applying the LED to the nodes inside each community (the former) and to coarse-grained communities (the latter). 
For this (literally) divide-and-conquer strategy, we use
the Louvain method~\cite{Blondel_JSM_Louvain,Reichardt_PRE_community} to systematically obtain community structures.
With our local-hub-based LED scheme applied to each community, we expect to identify core backbone nodes and peripheral shell nodes, which would be closely related to individual nodes' co-membership consistency with other nodes from stochastic community-detection algorithms (consistent core nodes versus inconsistent peripheral nodes)~\cite{HKim2019,DLee2021}.

First, we implement the LED in each community as shown in Fig.~\ref{fig6}(a); we treat each community as an individual network by only considering the nodes and edges inside. Again, we take the CB network~\cite{comgeo_net} and detect communities with the Louvain algorithm~\cite{Blondel_JSM_Louvain} using the resolution parameter~\cite{Reichardt_PRE_community} $\gamma = 0.5$ in the modularity function~\cite{Newman2004}. Although the resultant communities can be different for each realization by the algorithm's stochasticity~\cite{HKim2019,DLee2021}, we obtain $36$ communities in this case. Then, we take each community (except for $8$ communities with the topology of the star graph, where $S_{cp}$ is undefined because $D_{cc}$ is undefined~\footnote{Note that, in a sense, the star graph is the perfect CP and it should strongly positively contribute to $S_{cp}$.}) and decompose the nodes inside with the LED, by treating the community and connections inside as a network.
Since communities have low structural diversity inside, the distinction between the core and periphery is readily apparent, making it a suitable scale for identifying a single CP.
As an illustrative example with a particularly large $S_{cp}$ value, we take a community containing $575$ nodes with the maximum $S_{cp}(L_B^* = 9) = 0.378$.
Figure~\ref{fig6}(b) shows the connection densities and core-periphery score according to the boundary level, as in the entire network in Fig.~\ref{fig5}. 
To visually inspect the actual organization, we plot the community's adjacency matrix sorted by the nodes' hierarchy level in Fig.~\ref{fig6}(c).
In this case, the optimal value $L_B^* = 9$ corresponds to an intermediate decomposition stage.
The well-defined single CP inside a community is observed across most communities, and there are extreme CP cases where a single node forms the core (the star-graph structure) as discussed before.

To check the ``communities inside CP'' side, we take the coarse-grained point of view and treat each community detected by the Louvain algorithm~\cite{Blondel_JSM_Louvain,Reichardt_PRE_community} as a new node or a ``supernode'' for distinction. 
The supernode network is composed of supernodes and the edges connecting them if there is any connection between the members of each community, as illustrated in Fig.~\ref{fig7}(a) compared to Fig.~\ref{fig6}(a); multi-edges and self-loops are ignored for simplicity.
Then, on this coarse-grained network~\cite{Itzkovitz2005,Song2005}, we apply the same LED process to detect the CP. 
Figure~\ref{fig7}(b) shows the result for the supernode network from the CB network, again in the case of $\gamma = 0.5$.
The result indicates $L_B^* = 4$ in this case, which assigns the $7$-clique in the super-adjacency matrix depicted in Fig.~\ref{fig7}(c). 
The fact that $L_B^*$ corresponds to the final stage of decomposition may imply the lack of a characteristic scale or the lack of detailed resolution for this small-sized supernode network composed of $36$ supernodes; the latter hypothesis is supported by the same phenomenon happening for the LED inside a smaller community than the one depicted in Fig.~\ref{fig6}.
The super-adjacency matrix looks similar to the ideal CP form that we discussed;
the core nodes are densely connected and most periphery nodes are connected to core nodes.
As in the inside-community version of LED, the structural simplification performed by coarse-graining with Louvain communities enables us to find a simple CP. 
There are communities or supernodes playing a core role in the super-network composed of supernodes.
When using the Louvain method, the number of communities varies depending on the resolution parameter $\gamma$, and the nodes comprising each community change for each run even for a single $\gamma$ value due to its stochasticity.
However, except for extreme cases where the resolution is very close to $0$ or significantly greater than $1$, results are qualitatively similar to the result obtained in this section.

In addition, we remark on the utility of the hub centrality in this context. The degree as the number of neighbors measures a node's global significance within the entire network, which is certainly a useful and the most widely used one. However, depending on the purpose of the analysis, comparing the (global) degree values across different focused groups can make deciphering hidden organizational principles even harder, especially in the case of quite heterogeneous degree distributions characterizing real networks~\cite{Newman2018,Barabasi2016,MenczerFortunatoDavis2020}. The hub centrality~\cite{Jeong_2021,Jeong_2022} is a tailor-made measure to quantify each node's relative importance within each focused group composed of the node itself and its neighbors, and it can help us overcome such limitations by standardizing the status of local hubs (recall Fig.~\ref{fig:local_hub}). 
Utilizing the hub centrality, the LED assigns standardized levels to local units and identifies the CP based on these levels, and thus, it is versatile across different scales, such as a network, within a community, or a supernode network.

\subsection{LED versus \textit{k}CD for a synthetic network with a prescribed CP structure}\label{add2}\protect

\begin{figure}[tb]
\centering
\includegraphics[angle=0,width=0.7\columnwidth]{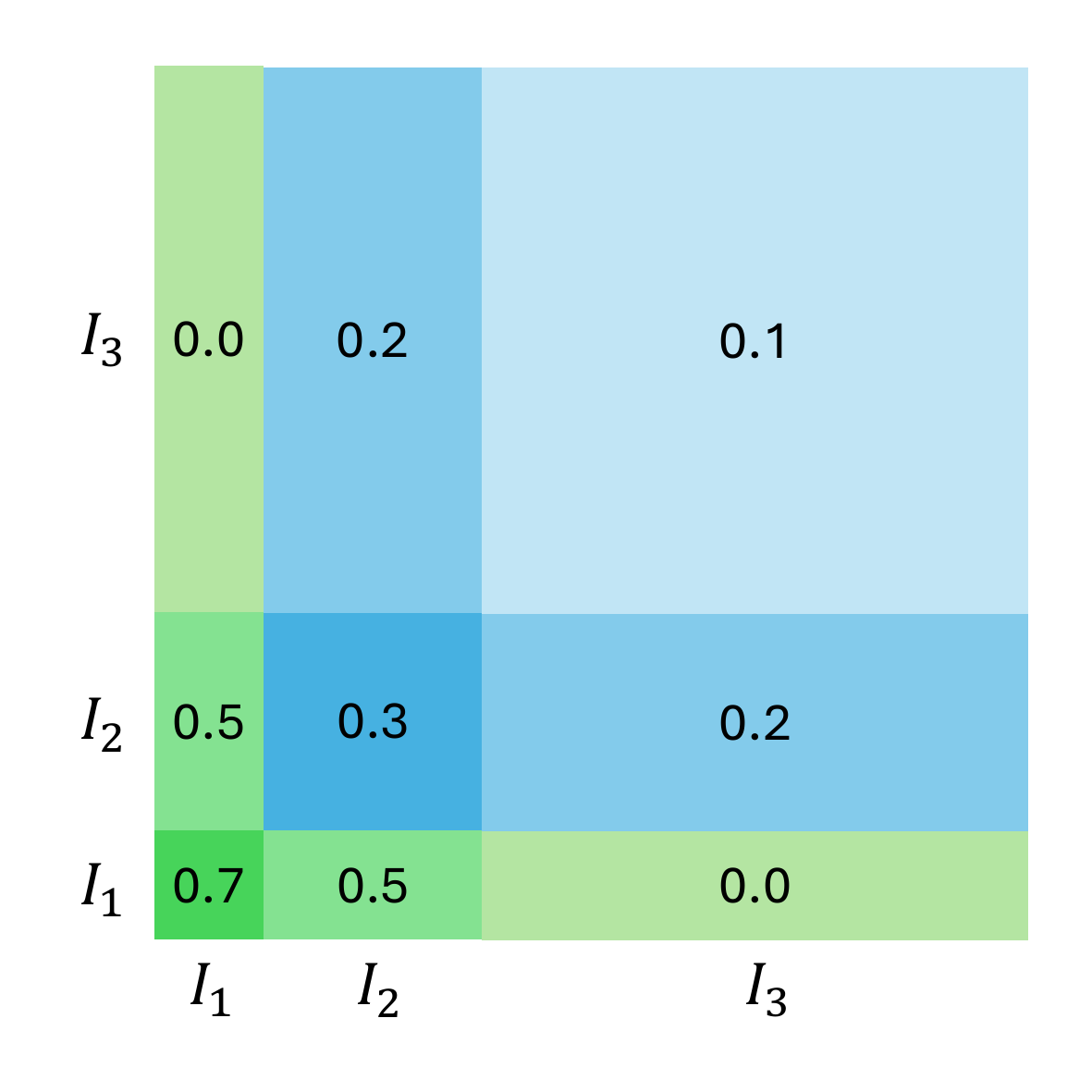}
\caption{%
Connection density scheme of the stochastic block model, which is composed of three parts: $I_1$, $I_2$, and $I_3$. The relative length of the side is proportional to the number of nodes in each block.
}%
\label{figEX}
\end{figure}

\begin{figure}[tb]
\centering
\includegraphics[angle=270,width=1\columnwidth]{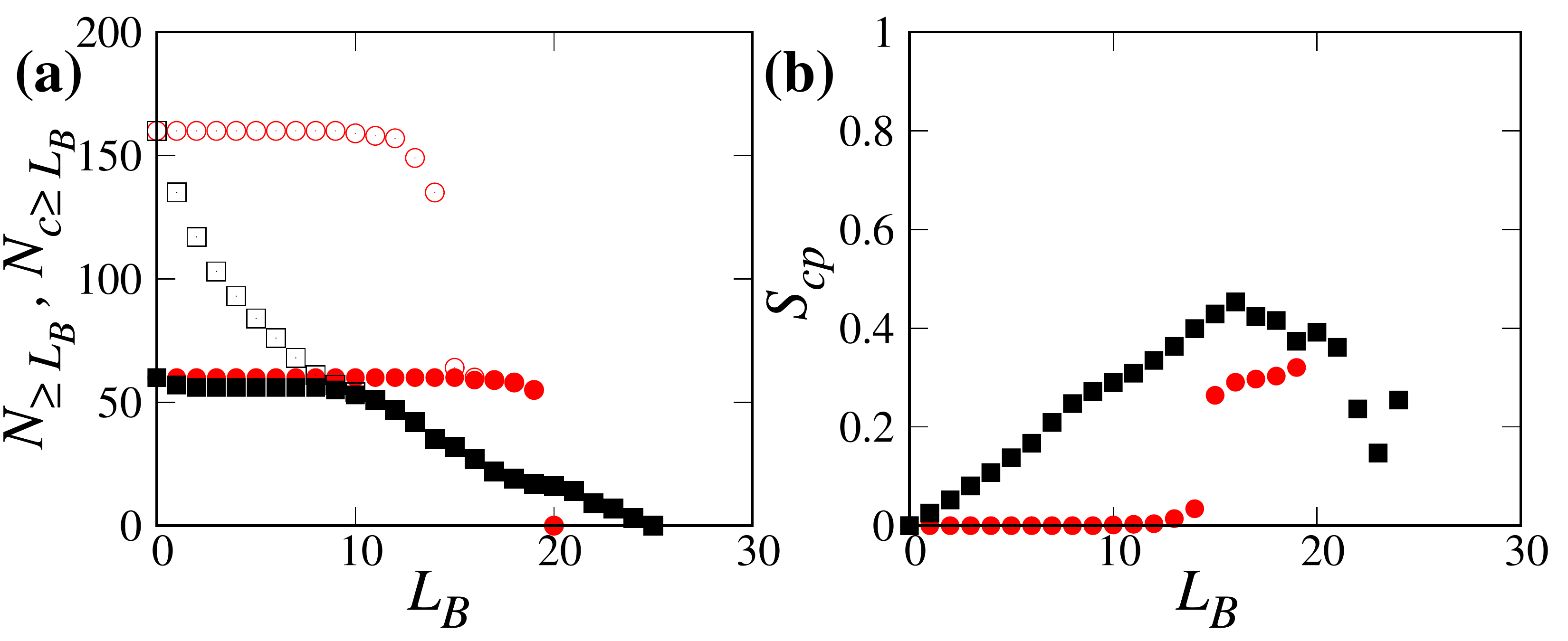}
\caption{%
(a) Number of core nodes and that of all nodes in the remained network with $L \ge L_B$ as a function of the boundary level $L_B$, from a single realization of the SBM structure in Sec.~\ref{add2}. The squares and circles represent the results for the LED and \textit{k}CD, respectively. The filled and open symbols indicate the core and all nodes, respectively. 
(b) Core-periphery score as a function of the boundary level. The filled squares and circles represent the cases of LED and \textit{k}CD, respectively.}%
\label{figSBM}
\end{figure}

In this final subsection, we demonstrate the LED's merit even for a simple structure with gradual coreness. To show a clear advantage of the LED over \textit{k}CD in this case, we generate a synthetic network using the stochastic block model (SBM)~\cite{Karrer2011}.
The synthetic network is composed of three parts denoted by $I_1$, $I_2$, and $I_3$, which are composed of $20$, $40$, and $100$ nodes, respectively.
As shown in Fig.~\ref{figEX}, the connection density between different parts is given by $D_{I_1I_1}=0.7$, $D_{I_2I_2}=0.3$, $D_{I_3I_3}=0.1$, $D_{I_1I_2}=0.5$, $D_{I_2I_3}=0.2$, and $D_{I_1I_3}=0$; the nodes inside each group and between groups are connected uniformly at random with the prescribed connection density. The structure represents the three-level coreness, where the primary core part $I_1$ do not have any connection to the periphery $I_3$ and the secondary core $I_2$ plays the role of bridge between the primary core and the periphery. 
As shown in Fig.~\ref{figSBM}(a), in the case of \textit{k}CD-based decomposition, the number of nodes with $L \ge L_B$ in the network does not notably change when $L_B < 10$. In other words, even the level of most peripheral nodes typically goes up to higher than $L_B = 10$, and only the $60$ nodes constituting the primary and secondary cores remain when $L_B \ge 16$. It means that the \textit{k}CD roughly cuts the structure into two levels: both the primary and the secondary cores as the core and the rest as the periphery. 
When we apply the LED, in sharp contrast, the number of total nodes with $L \ge L_B$ decreases gradually, and in particular, the number of primary and secondary core nodes ($60$ nodes) gradually decreases for $L_B > 10$, while most decrement in the number of nodes for $L_B < 10$ is caused by the removal of peripheral nodes ($100$ nodes). Even if we set up the three-level coreness, due to the statistical fluctuation there must be differential levels of local hubs for a single realization of SBM, and the LED successfully captures them as well. 

We also measure $S_{cp}$ as a function of $L_B$, in the same manner as in Fig.~\ref{fig5}.
As shown in Fig.~\ref{figSBM}(b), in the case of the \textit{k}CD, since most periphery nodes are in high levels ($> 10$), $S_{cp}$ remains low across a wide range of lower $L_B$ values.
At the maximum value of $L_B = 19$, $S_{cp}$ reaches $0.320$ and the most densely connected $55$ nodes are classified as the core.
However, in the case of the LED, $S_{cp}$ starts to increase and reaches its maximum value ($0.453$) at $L_B = L_B^* = 16$, and decreases from that point. The number of nodes at $L_B^*$ is $20$ [Fig.~\ref{figSBM}(a)], which corresponds to the primary core nodes as expected. In other words, when using the LED, it is possible to identify the most influential nodes even within densely connected nodes. 
This behavior is consistent with that observed in real networks, shown in Figs.~\ref{fig5} and \ref{fig6}.
As $L_B$ increases, the proportion of nodes belonging to $I_1$ increases among the nodes with $L > L_B$, leading to a reduction in the number of connections between nodes with $L \geq L_B$ and those with $L < L_B$.
This results in a lower value of $N_{cp}/N_p$ and the presence of the optimal boundary level. 
Compared to the \textit{k}CD, the changes in $S_{cp}$ values are smoother in the LED. In the real-world networks considered, many communities exhibit star or densely connected structures. In those cases, the \textit{k}CD often assigns a single level to all nodes, effectively grouping them at the same level. In contrast, the LED differentiates nodes by assigning various levels on them, which naturally leads to detecting an appropriate core in that structures. Consequently, the LED has clear advantages over \textit{k}CD in identifying a proper cores in these specific structures.

\section{Summary and Outlook}\label{sec:level4}\protect

The most important insight from our analysis is the possibility of using the concept of local hubs, defined with respect to the degrees of the neighbors of a node of interest. We believe that the local hubs can be seeds (potentially corresponding to the local cores in terms of core-periphery separation) for a different type of network decomposition, compared with conventional ones such as the $k$-core decomposition from the global perspective.
For that purpose, we have proposed a method to uncover the core-periphery structure of networks through network decomposition centered around local hubs. Compared with other edge or node centralities such as edge betweenness and degree-product, our hub-centrality-product rule ensures the existence of a series of natural cutoffs in the form of a cusp on the giant component size versus the fraction of removed edges. The cusp point, unambiguously defined as the transition point of removed edges with zero versus nonzero hub-centrality product, signals the breaking point of a shell from a backbone. 
Our local-edge decomposition method repeats this process for each backbone as a brand-new network until we run out of edges to be removed. 
We have demonstrated the properties and implications of the method with a collaboration network as a representative example, in particular, compared with the celebrated \textit{k}-core decomposition. 

Naturally, the decomposition yields the core-periphery separation and we have introduced a principled way to pinpoint the boundary by the core-periphery score accounting the similarity to the ideal core-periphery structure with respect to the background edge density; in particular, we emphasize again that our local-edge decomposition shines when it is combined with community division, by preparing the groups with appropriate sizes to handle with the decomposition in advance.
We believe that the combination leads to the most natural way to extract so-called multiple core-periphery structures~\cite{Kojaku_PRE_cp,Kojaku_NJP_cp,Yang_ESA_cp,Gallagher_SCIADV_cp}, and the extension to the coarse-grained communities as supernodes clearly demonstrates the intermingled structure of core-periphery and communities~\cite{Rombach2017}.
From the result, we have clearly shown the merit of embracing local hubs in structurally dissecting networks, as in the previously reported effect on dynamical properties~\cite{Jeong_2021,Jeong_2022}. 
Some related previous works by others include Ref.~\cite{email_net} on community structures, where so-called ``whiskers'' can be recognized as part of the periphery despite having relatively high degrees because it is based on relative connections. This approach may align in many aspects with the study that utilizes the hub centrality.
Moreover, a comprehensive comparison between our approaches (LED and its application to detect core-periphery structures) and existing methods~\cite{Borgatti2000,Holme2005,Csermely2013,Rombach2017,Tang_CCF_cp,Kojaku_PRE_cp,Kojaku_NJP_cp,Yang_ESA_cp,Gallagher_SCIADV_cp} will provide valuable insights in the core-periphery structure, and we leave it as foreseeable and plausible future research.
Most of all, we hope that this type of perspective regarding locally important substructures, such as the concept of hidden dependency between nodes from it~\cite{MJLee2021}, gets more attention from the network science community.

\section*{Acknowledgments}\protect
This work was supported by the National Research Foundation (NRF) of Korea under Grant No.~RS-2021-NR061247. The authors thank Ludvig Lizana for the discussion regarding the concept of using supernodes from network communities in a coarse-grained level, Heetae Kim for insightful comments including correcting typos, and an anonymous referee for valuable suggestions for future work.

\end{CJK}

\bibliography{References}% Produces the bibliography via BibTeX.

\end{document}